\documentclass[prd,aps,superscriptaddress,nofootinbib,showpacs,twocolumn]{revtex4}

\usepackage[intlimits]{amsmath}
\usepackage{amsfonts}
\usepackage{amssymb,amscd}
\usepackage{color}
\usepackage[dvips]{epsfig}

\newcommand{\beq}{\begin{equation}}
\newcommand{\eeq}{\end{equation}}
\newcommand{\beqa}{\begin{eqnarray}}
\newcommand{\eeqa}{\end{eqnarray}}
\newcommand{\be}{\begin{equation}}
\newcommand{\ee}{\end{equation}}

\newcommand{\vp}{\vec{p}}
\newcommand{\vq}{\vec{q}}
\newcommand{\vk}{\vec{k}}
\newcommand{\op}{\omega_{p}}
\newcommand{\oq}{\omega_{q}}
\newcommand{\ok}{\omega_{k}}

\newcommand{\gS}[1]{#1\!\!\!\!\!\not~}

\newcommand{\pslash}{\gS{p}}

\def\eq#1{(\ref{#1})}
\def\Eq#1{Eq.~(\ref{#1})}

\begin{document}

\title{Deconfinement phase transition and the quark condensate}

\author{Christian~S.~Fischer}
\affiliation{Institut f\"ur Kernphysik, 
  Technische Universit\"at Darmstadt,
  Schlossgartenstra{\ss}e 9,\\ 
  D-64289 Darmstadt, Germany}
\affiliation{GSI Helmholtzzentrum f\"ur Schwerionenforschung GmbH, 
  Planckstr. 1  D-64291 Darmstadt, Germany.}

\date{\today}
\begin{abstract}
We study the dual quark condensate as a signal for 
the confinement-deconfinement phase transition of QCD.
This order parameter for center symmetry has been defined 
recently by Bilgici et al. within the framework of lattice 
QCD. In this work we determine the ordinary and the dual 
quark condensate with functional methods using a formulation 
of the Dyson-Schwinger equations for the quark propagator 
on a torus. The temperature dependence of these condensates 
serves to investigate the interplay between the chiral and 
deconfinement transitions of quenched QCD. 
\end{abstract}

\maketitle

{\bf Introduction}\\ 
The chiral and deconfinement transition of QCD is a 
subject of continuous interest. One of the unsolved 
problems is the question of an underlying mechanism 
relating both phenomena. Strictly speaking chiral 
and deconfinement phase transitions only occur in 
opposite sectors of the theory. The chiral condensate 
acts as order parameter for the chiral phase transition 
at vanishing quark masses, $m=0$, whereas the Polyakov 
loop signals center symmetry breaking at the deconfinement 
transition for $m \rightarrow \infty$. At intermediate 
masses both transitions develop a fascinating interplay 
\cite{Karsch:1998ua}, which is not yet understood in 
detail. Nonperturbative methods such as lattice gauge 
theory or functional methods are needed to explore this 
issue. 

A recent development that sheds light on this connection 
is the investigation of spectral sums of the Dirac 
propagator and their behaviour under center transformations. 
Initiated by a work of Gattringer \cite{Gattringer:2006ci} 
spectral sums have been explored in 
Refs.~\cite{Bruckmann:2006kx,Synatschke:2007bz,%
Bilgici:2008qy,Synatschke:2008yt}.
In particular it has been shown that the quark confinement 
mechanism is entirely encoded in the low lying spectral 
modes of the Dirac operator \cite{Synatschke:2008yt},
which are also responsible for chiral symmetry breaking via
the celebrated Casher-Banks relation. The spectral sums
constitute order parameters for the deconfinement 
transition of QCD \cite{Gattringer:2006ci,Synatschke:2008yt}. 

In general, order parameters for deconfinement are not 
easily accessible by functional methods. In \cite{Braun:2007bx}
a method has been developed to determine the Polyakov loop 
potential from the effective action, whereas in 
\cite{Bender:1996bm} the analytic structure of the quark 
propagator has been used to distinguish between the confined and
deconfined phase. In this letter we report on a 
calculation of another order parameter for deconfinement,
the dual quark condensate or 'dressed Polyakov 
loop' \cite{Bilgici:2008qy}, with functional methods.

This dual condensate $\Sigma_1$ is 
defined by the Fourier-transform
\beq \label{dual}
\Sigma_1 = -\int_0^{2\pi} \, \frac{d \varphi}{2\pi} \, e^{-i\varphi}\, 
\langle \overline{\psi} \psi \rangle_\varphi
\eeq
of the ordinary quark condensate 
$\langle \overline{\psi} \psi \rangle_\varphi$ evaluated using 
$U(1)$-valued boundary conditions with angle $\varphi$ in the 
temporal direction of Euclidean, quenched QCD. Note that the usual
antiperiodic boundary conditions for fermions require $\varphi=\pi$, 
whereas $\varphi=0$ corresponds to periodic boundary conditions;
here we vary $\varphi$ in the interval $[0,2\pi]$. 
In order to explain why the quantity $\Sigma_1$ is of considerable 
interest we note that the $\varphi$-dependent quark condensate 
$\langle \overline{\psi} \psi \rangle_\varphi$ can be represented 
by a sum over all possible closed chains of link variables, 
i.e. closed loops $l$, on a lattice. One obtains:
\beq \label{loop}
\langle \overline{\psi} \psi \rangle_\varphi =  
\sum_{l} \frac{e^{i\varphi n(l)}}{m^{|l|}} U(l) \,,
\eeq
where $U(l)$ denotes the closed chains of links including some sign and
normalisation factors, see \cite{Bilgici:2008qy} for details. Each
loop consists of $|l|$ links and is weighted with the corresponding
power of the inverse quark mass $m$. Each time such a closed loop winds 
around the temporal direction of the lattice it picks up a factor 
$e^{\pm i\varphi}$ from the $U(1)$-valued boundary condition introduced 
above. This leads to a weighting $e^{i\varphi n(l)}$, where $n(l)$
is the winding number of a given loop $l$. The
dressed Polyakov loop $\Sigma_1$, Eq.~(\ref{dual}), projects out loops
with $n(l)=1$. It transforms under center 
transformation in the same way as the conventional Polyakov loop 
\cite{Polyakov:1978vu} and is therefore an order parameter for 
the deconfinement transition. The numerical agreement between 
dressed and conventional Polyakov loop has been established for 
gauge groups $SU(3)$ \cite{Bruckmann:2008sy} and, remarkably, also 
for the centerless $G(2)$ \cite{Danzer:2008bk}. 

In this work we determine the dual quark condensate from the 
Dyson-Schwinger equations (DSEs) of Landau gauge QCD at finite 
temperature \cite{Roberts:2000aa}. 
However, we wish to point out that our method is sufficiently 
general to be of equal use in other functional approaches as 
e.g. the functional renormalisation group \cite{Braun:2006jd}. 
We evaluate the ordinary and the dual quark condensate from the 
trace of the quark propagator in a formulation of the DSEs 
on a torus. Our investigation of chiral symmetry breaking 
and deconfinement complements corresponding ones at vanishing 
temperature, see e.g. \cite{Alkofer:2008tt}, and provides an interesting tool 
for further investigations of the QCD phase diagram by functional 
methods.

{\bf Dyson-Schwinger equations on a torus}\\
We work in Euclidean space with compact time and space
directions, i.e. box size $V=L^3 \times 1/T$ with temperature $T$ and 
$1/T << L$. We choose periodic boundary conditions 
in the three spatial directions for the quark and gluon fields. The 
gluon field also obeys periodic boundary conditions in the temporal 
direction. For the fermion field $\psi$ we use the generalised, 
$U(1)$-valued boundary condition 
$\psi(\vec{x},1/T) = e^{i \varphi} \psi(\vec{x},0)$ as described above. 
In loop integrals in momentum space this results in Matsubara 
modes $\op(n_t) = (2\pi T)(n_t+\varphi/2\pi)$ in the $p_4$-direction, 
which depend on the boundary angle $\varphi \in [0,2\pi[$. In the 
spatial directions we have the usual Matsubara sums which are treated
with the techniques described in \cite{Fischer:2002eq}.

\begin{figure}[t]
\centerline{\includegraphics[width=\columnwidth]{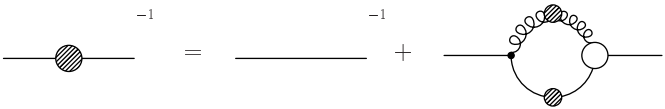}}
\caption{The Dyson-Schwinger equation for the quark propagator. Filled circles
denote dressed propagators whereas the empty circle stands for the dressed
quark-gluon vertex.}
\label{fig:quarkDSE}
\end{figure}
In Euclidean momentum space at nonzero temperature the renormalised dressed 
quark propagator is given by
\beq \label{quark}
S(\vp,\op) = [i \gamma_4 \op C(\vp,\op) + i \vec{\pslash} A(\vp,\op) + B(\vp,\op)]^{-1} \,,
\eeq
with vector and scalar quark dressing functions $C,A,B$. For the  
bare quark propagator $S_0$ we have $A=C=1$ and $B= Z_m m$, 
with bare quark mass $m$. The DSE for the quark propagator,
shown in Fig.~\ref{fig:quarkDSE}, reads
\begin{widetext}
\beqa \label{DSE}
S^{-1}(\vp,\op) = Z_2 \, S^{-1}_0(\vp,\op) 
-  C_F\, \frac{Z_2}{\widetilde{Z}_3}\, \frac{g^2 T}{L^3} \sum_{n_t,n_i} \,
\gamma_{\mu}\, S(\vk,\ok) \,\Gamma_\nu(\vk,\ok,\vp,\op) \,
D_{\mu \nu}(\vp-\vk,\op-\ok) \,. \label{quark_t}
\eeqa
\end{widetext}
where the sum is over temporal and spatial Matsubara modes.
The Casimir $C_F = (N_c^2-1)/N_c$ stems from the colour trace; 
in this work we only consider the gauge group $SU(2)$.
Furthermore, $D_{\mu \nu}$ denotes the gluon propagator in Landau gauge and 
$\Gamma_\nu$ the (reduced) quark-gluon vertex. 
The renormalisation factors $Z_2$, $Z_m$ and 
$\widetilde{Z}_3$ are determined in the renormalisation process. 

\begin{table}[b]
\begin{tabular}{c||c|c|c|c}
T[MeV]     & 0 & 119  & 298  & 597  \\\hline\hline
$a_{T}(T)$   & 1 & 1    & 1.34 & 1.65 \\   
$a_{L}(L)$   & 1 & 1    & 0.8  & 4.0    
\end{tabular}
\caption{Temperature dependent fit parameter of Eq.(\ref{gluefit}).\label{tab}}
\end{table}
In order to solve this equation we have to specify explicit expressions
for the gluon propagator and the quark-gluon vertex. For the momentum range
relevant for \Eq{DSE} we nowadays have very accurate solutions for the gluon
propagator at zero temperature from both, lattice calculations and functional
methods, see e.g. \cite{Fischer:2008uz} and references therein. The 
temperature dependence of the gluon propagator, however, is much less 
explored. In \cite{Cucchieri:2007ta} a combined lattice and DSE study records
a different temperature dependence of the electric and magnetic
sector. Whereas the magnetic part of the propagator seems to be indifferent 
to the deconfinement phase transition, the electric part is strongly
increased at and around the critical temperature $T_c \approx 300$ MeV.
Although the
lattice data still have considerable systematic errors they may 
correctly represent the qualitative temperature dependence of the gluon. 
We therefore use a temperature dependent fit to the data given by

\begin{eqnarray}
D_{\mu\nu}(q) = \frac{Z_T(q)}{q^2} P_{\mu \nu}^T(q) 
                    +\frac{Z_L(q)}{q^2} P_{\mu \nu}^L(q) 
\end{eqnarray} 
with $q=(\vq,\oq)$ and dressing functions $Z_T(\vq,\oq)$ and $Z_L(\vq,\oq)$. 
The transverse and longitudinal projectors with
respect to the heat bath are given by 
\begin{eqnarray}
P_{\mu\nu}^T(q) &=& 
   \left(\delta_{i j}-\frac{q_i q_j}{\vq^2}\right) 
   \delta_{i\mu}\delta_{j\nu}\,, \nonumber\\
P_{\mu\nu}^L(q) &=& P_{\mu \nu}(q) - P_{\mu \nu}^T(q) \,,
\end{eqnarray} 
with ($i,j=1 \dots 3$). 
The $SU(2)$ lattice results of Ref.~\cite{Cucchieri:2007ta} are well fitted by  
\beqa \label{gluefit}
Z_{T,L}(\vq,\oq,T) &=& \frac{q^2 \Lambda^2}{(q^2+\Lambda^2)^2} \,
\left\{\left(\frac{c}{q^2+ \Lambda^2 a_{T,L}(T)}\right)^2 \right.\nonumber\\
&&\left.+\frac{q^2}{\Lambda^2}\left(\frac{\beta_0 \alpha(\mu)\ln[q^2/\Lambda^2+1]}{4\pi}\right)^\gamma\right\}
\eeqa
with the temperature independent scale $\Lambda = 1.4$ GeV and 
the coefficient $c=9.8 \,\mbox{GeV}^2$. For gauge group $SU(2)$ we have
$\beta_0 = 22/3$ and $\gamma=-13/22$ in the quenched theory and we renormalise 
at $\alpha(\mu)=0.3$. The temperature dependent 
scale modification parameters $a_{T,L}(T)$ are given in table \ref{tab}.
In order to extend this fit to temperatures not given in the table we assume 
$a_{T,L}(T)$ to be temperature independent below $T=119$ MeV and only slowly 
rising above $T=597$ MeV. For $T \in [119,597]$ MeV we use cubic splines to 
interpolate smoothly between the values given in table \ref{tab}. We expect 
the systematic error of this procedure to be of the same order as the 
systematic errors inherent in the lattice data. We also inherit the scale 
determined on the lattice using the string tension 
$\sqrt{\sigma}=0.44$ GeV \cite{Cucchieri:2007ta}.  

\newpage
For the quark-gluon vertex with gluon momentum $q=(\vq,\oq)$ and the quark 
momenta $p=(\vp,\op),k=(\vk,\ok)$ we employ the following temperature 
dependent model
\begin{widetext}
\beqa \label{vertexfit}
\Gamma_\nu(q,k,p) = \widetilde{Z}_{3}\left(\delta_{4 \nu} \gamma_4 
\frac{C(k)+C(p)}{2}
+  \delta_{j \nu} \gamma_j 
\frac{A(k)+A(p)}{2}
\right)\left( 							
\frac{d_1}{d_2+q^2} 			
 + \frac{q^2}{\Lambda^2+q^2}
\left(\frac{\beta_0 \alpha(\mu)\ln[q^2/\Lambda^2+1]}{4\pi}\right)^{2\delta}\right) \,.
\eeqa 
\end{widetext}
where $\delta=-9/44$ is the anomalous dimension of the vertex. Note that because of
$\gamma+2\delta=-1$ the gluon dressing function together with the quark-gluon vertex 
behave like a running coupling at large 
momenta; this is a necessary boundary condition for any model 
interaction in the quark DSE. The dependence of the vertex on the quark dressing
functions $A$ and $C$ is dictated by the Slavnov-Taylor identity. The remaining 
fit function is purely
phenomenological, see e.g. \cite{Fischer:2008wy} where an elaborate 
version of such an ansatz has been used to describe meson observables. 
Here we use $d_1 = 7.6 \,\mbox{GeV}^2$ and $d_2=0.5 \,\mbox{GeV}^2$.
A variation of these parameters shifts the critical temperatures of both, 
the chiral and the deconfinement transition but leaves all qualitative 
aspects of the results presented below unchanged. This is also true if
one changes the deep infrared behaviour of the vertex to the (infrared 
divergent) form extracted in \cite{Alkofer:2008tt} and used in 
\cite{Fischer:2008wy}; the confinement-deconfinement phase transition 
is insensitive to the question of scaling vs. 
decoupling in the sense specified in \cite{Fischer:2008uz}. 
Details will be reported elsewhere. 

The DSE is solved numerically using $n_t=8$ Matsubara 
frequencies and a three-volume of $V=(5 \,\mbox{fm})^3$, which is
large enough to avoid any significant volume effects.
The renormalisation conditions are 
$C(\mu) = 1$ and $B(\mu) = m$ with $\mu=(\vec{\mu},\pi T)$ and 
$\vec{\mu}^2=20 \,\mbox{GeV}^2$.
The resulting quark dressing functions are subsequently 
used to determine the quark condensate according to
\beq
\langle \overline{\psi} \psi \rangle_\varphi =   \frac{4 Z_2 N_c T}{L^3} \sum_{n_t,n_i}
\frac{B(\vp,\op(\varphi))}{\op^2(\varphi) C^2 + \vec{p}^2 A^2 + B^2}\,,
\eeq
where we indicated the dependence of the frequencies on the generalised 
$U(1)$-boundary conditions. For nonvanishing bare quark masses $m$ this 
expression is divergent and, at least in the continuum limit, has to be 
renormalised accordingly. For the purpose of this letter, however, is 
is sufficient to work with the regularised expression at fixed ultraviolet
cut-off corresponding to fixed lattice spacing.

{\bf Numerical results}\\
\begin{figure}[b]
\centerline{\includegraphics[width=0.9\columnwidth]{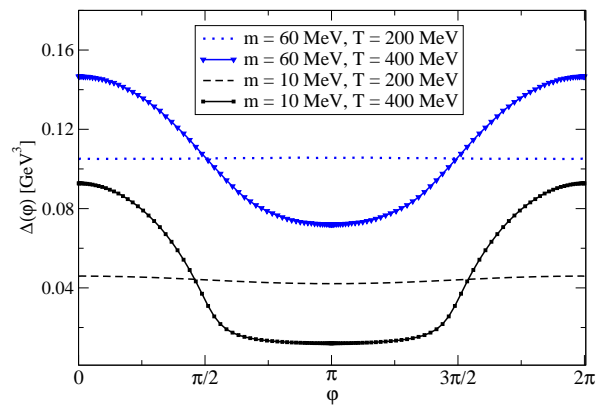}}
\caption{The angular dependence of the quark condensate 
$\Delta_\varphi \equiv \langle \overline{\psi} \psi \rangle_\varphi$ 
below and above the deconfinement transition
for two different quark masses.}
\label{fig:pic3}
\end{figure}
First we explore the dependence of the quark condensate on the boundary 
angle $\varphi$ which can be read off \Eq{loop}. Each loop winding $n$ 
times around the temporal
direction of the torus contributes a factor $\cos(n\varphi)$. Consequently
the integrand in \Eq{dual} can be expanded as a series in $\cos(n\varphi)$
and is symmetric in the interval $[0,2\pi]$. This behaviour is clearly seen
also in our numerical result for 
$\Delta_\varphi \equiv \langle \overline{\psi} \psi \rangle_\varphi$
shown in Fig.\ref{fig:pic3}. For $T=200$ MeV far below the deconfinement
transition we find almost no angular dependence of the condensate. This 
is especially true for the heavier quark mass $m = 60 \,\mbox{MeV}$.
For $T=400$ MeV far above the transition the behaviour is markedly different 
and we observe the characteristic cosine-type of behaviour of the
condensate that we expect from \Eq{loop}. This is in nice agreement with the 
results of Ref.~\cite{Bilgici:2008qy} on the lattice.
The angular dependence of the condensate can be fitted well with a series 
in $\cos(n\varphi)$. Whereas for the heavier mass $m = 60 \,\mbox{MeV}$
terms with $n \le 3$ are sufficient one needs at least terms up to $n \le 7$ for the 
smaller mass $m = 10 \,\mbox{MeV}$. Thus the closer one gets to 
the chiral limit the more contributions from Polyakov loops with higher
winding number around the temporal direction are significant. This is a 
direct consequence of the mass factor $1/m^{|l|}$ in the expansion \Eq{loop}
and also seen on the lattice \cite{private}. 
\begin{figure}[t]
\centerline{\includegraphics[width=0.88\columnwidth]{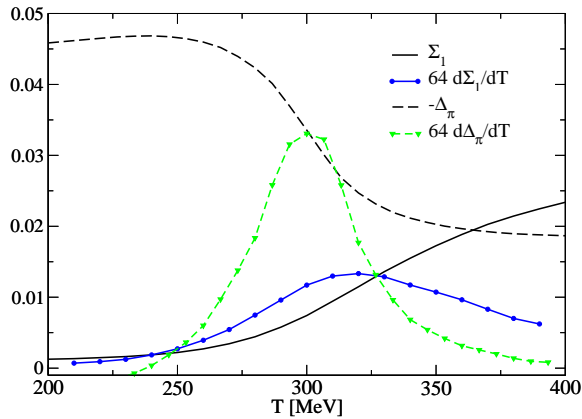}}
\caption{The temperature dependence of the dressed Polyakov-loop 
$\Sigma_1$
and the conventional quark condensate 
$\Delta_\pi \equiv \langle \overline{\psi} \psi \rangle_{\varphi=\pi}$  
together with their derivatives for $m = 10 \,\mbox{MeV}$.}
\label{fig:pic12}
\end{figure}

By far the largest contribution to the angular dependence of the condensate 
comes from loops with $n=1$ which are
projected out by the Fourier transformation \eq{dual} to the dual
condensate. The resulting temperature dependence of the 
dressed Polyakov loop is shown in Fig.~\ref{fig:pic12} 
together with the conventional quark condensate and their derivatives.
One clearly observes a change of behaviour in both, the conventional and 
the dual condensate above $T=270$ MeV. The temperature derivative of both 
quantities has a peak in the region $T_c=300-320$ MeV signalling the 
chiral and deconfinement transition. Note that although both transitions 
are calculated from quantities with direct relation to the spectral 
properties of the quark propagator they do not necessarily give the same
transition temperatures. On the contrary, the chiral transition occurs
about $10-20$ MeV below the deconfinement transition. Whether the quantitative 
aspects of this difference is a model-independent result has to be 
investigated in more detail. 

Finally we wish to point out that the absolute magnitude of our dressed 
Polyakov loop is about a factor of five smaller than the one calculated 
on the lattice \cite{Bilgici:2008qy}. Since the condensate and subsequently 
its dual is a renormalisation point dependent quantity we attribute this 
difference to a yet undetermined renormalisation factor. This issue has to 
be clarified in future work. Furthermore we note that the dressed Polykov 
loop is not strictly zero in the low temperature region as expected in 
quenched QCD. This is certainly a consequence of our vertex ansatz which
does not strictly represent quenched QCD but includes a (small) amount
of unquenching effects.

{\bf Concluding remarks}\\
In this letter we have determined the dual condensate or dressed Polyakov 
loop by solving the Dyson-Schwinger equations for the quark propagator
on a compact manifold. This order parameter for center symmetry measures
contributions from closed loops winding once around the time direction.
We observe a significant rise of the dressed Polyakov loop around and above
the deconfinement transition temperature together with a significant
decrease of the ordinary quark condensate. The angular dependence of the
quark condensate shows a characteristic dependence of the quark mass: the
lighter the quark the more contributions arise from loops with winding
numbers larger than one. An obvious next step is to investigate what 
happens to the transition temperatures when the backreaction of the 
quarks onto the Yang-Mills sector is included. Particularly interesting 
in this respect is the case $N_f=2+1$ which recently is a subject of 
intense debate \cite{Aoki:2009sc,Bazavov:2009zn} in the lattice community.

{\bf Acknowledgments}\\
I thank Falk Bruckmann, Christoph Gattringer, Jens M\"uller and 
Jan Pawlowski for discussions. I am grateful to Axel Maas 
for discussions and for making the lattice data of Ref.~\cite{Cucchieri:2007ta} 
available. This work has been supported by the
Helmholtz-University Young Investigator Grant number VH-NG-332.\\[1mm]


\begin{thebibliography}{99}

\bibitem{Karsch:1998ua}
  F.~Karsch,
  arXiv:hep-lat/9903031.

\bibitem{Gattringer:2006ci}
  C.~Gattringer,
  Phys.\ Rev.\ Lett.\  {\bf 97} (2006) 032003.

\bibitem{Bruckmann:2006kx}
  F.~Bruckmann, C.~Gattringer and C.~Hagen,
  Phys.\ Lett.\  B {\bf 647} (2007) 56.

\bibitem{Synatschke:2007bz}
  F.~Synatschke, A.~Wipf and C.~Wozar,
  Phys.\ Rev.\  D {\bf 75} (2007) 114003;


\bibitem{Bilgici:2008qy}
  E.~Bilgici, F.~Bruckmann, C.~Gattringer and C.~Hagen,
  Phys.\ Rev.\  D {\bf 77} (2008) 094007.

\bibitem{Synatschke:2008yt}
  F.~Synatschke, A.~Wipf and K.~Langfeld,
  Phys.\ Rev.\  D {\bf 77} (2008) 114018.

\bibitem{Braun:2007bx}
  J.~Braun, H.~Gies and J.~M.~Pawlowski,
  arXiv:0708.2413 [hep-th];
  F.~Marhauser and J.~M.~Pawlowski,
  arXiv:0812.1144 [hep-ph].

\bibitem{Bender:1996bm}
  A.~Bender, D.~Blaschke, Y.~Kalinovsky and C.~D.~Roberts,
  Phys.\ Rev.\ Lett.\  {\bf 77} (1996) 3724.


\bibitem{Polyakov:1978vu}
  A.~M.~Polyakov,
  Phys.\ Lett.\  B {\bf 72} (1978) 477;
  L.~Susskind,
  Phys.\ Rev.\  D {\bf 20}, 2610 (1979).

\bibitem{Bruckmann:2008sy}
  F.~Bruckmann, C.~Hagen, E.~Bilgici and C.~Gattringer,
  PoS {\bf LATTICE2008} (2008) 262.

\bibitem{Danzer:2008bk}
  J.~Danzer, C.~Gattringer and A.~Maas,
  JHEP {\bf 0901} (2009) 024.

\bibitem{Roberts:2000aa}
  C.~D.~Roberts and S.~M.~Schmidt,
  Prog.\ Part.\ Nucl.\ Phys.\  {\bf 45}, S1 (2000);
  A.~Maas, J.~Wambach and R.~Alkofer,
  Eur.\ Phys.\ J.\  C {\bf 42} (2005) 93.

\bibitem{Braun:2006jd}
  J.~Braun and H.~Gies,
  JHEP {\bf 0606} (2006) 024.

\bibitem{Alkofer:2008tt}
  R.~Alkofer, C.~S.~Fischer, F.~J.~Llanes-Estrada and K.~Schwenzer,
  Annals Phys.\  {\bf 324} (2009) 106.


\bibitem{Fischer:2002eq}
  C.~S.~Fischer, R.~Alkofer and H.~Reinhardt,
  Phys.\ Rev.\  D {\bf 65} (2002) 094008;
  C.~S.~Fischer and M.~R.~Pennington,
  Phys.\ Rev.\  D {\bf 73} (2006) 034029;
  C.~S.~Fischer, A.~Maas, J.~M.~Pawlowski and L.~von Smekal,
  Annals Phys.\  {\bf 322} (2007) 2916.

\bibitem{Fischer:2008uz}
  C.~S.~Fischer, A.~Maas and J.~M.~Pawlowski,
  arXiv:0810.1987 [hep-ph].

\bibitem{Cucchieri:2007ta}
  A.~Cucchieri, A.~Maas and T.~Mendes,
  Phys.\ Rev.\  D {\bf 75}, 076003 (2007).

\bibitem{Fischer:2008wy}
  C.~S.~Fischer and R.~Williams,
  Phys.\ Rev.\  D {\bf 78} (2008) 074006.


\bibitem{private}
 F.~Bruckmann, C.~Gattringer and C.~Hagen, private communication.


\bibitem{Aoki:2009sc}
  Y.~Aoki {\it et al.}, 
  arXiv:0903.4155 [hep-lat].

\bibitem{Bazavov:2009zn}
  A.~Bazavov {\it et al.},
  arXiv:0903.4379 [hep-lat].


\end{thebibliography}
\end{document}